# Analytic Comparison between X-ray Fluorescence CT and K-edge CT

Peng Feng, Ge Wang, *Fellow*, *IEEE*, Wenxiang Cong, Biao Wei

*Abstract*—X-ray fluorescence computed tomography (XFCT) and K-edge computed tomography (CT) are two important modalities to quantify a distribution of gold nanoparticles (GNPs) in a small animal for preclinical studies. It is valuable to determine which modality is more efficient for a given application. In this paper, we report a theoretical analysis in terms of signal-to-noise ratio (SNR) for the two modalities, showing that there is a threshold for the GNPs concentration and XFCT has a better SNR than K-edge CT if GNPs concentration is less than this threshold. Numerical simulations are performed and two kinds of phantoms are used to represent multiple concentration levels and feature sizes. Experimental results illustrate that XFCT is superior to K-edge CT when contrast concentration is lower than 0.4% which coincides with the theoretical analysis.

*Index Terms*—Contrast-to-noise ratio (CNR), gold nanoparticles (GNPs), image noise, image resolution, K-edge CT, x-ray fluorescence computed tomography (XFCT).

## I. INTRODUCTION

X-RAY computed tomography (CT), including K-edge CT, has an important role in diagnosis, staging, treatment planning, and often as an unique imaging examination method in hospitals and clinics[1]-[3]. Over the past decade, x-ray micro-CT has also seen a great development for small animal imaging, which is a key part in research of phenotyping, drug and disease mechanisms. While x-ray fluorescence computed tomography (XFCT) with synchrotron radiation used to be irrelevant to biomedical applications, it has recently attracted a major attention for preclinical research[4][5]. XFCT can be seen as a stimulated emission tomography, in which a sample is irradiated with x-rays more energetic than the K-shell energy of the target elements of interest. This will produce fluorescence x-rays isotropically emitted from the sample, and the characteristic x-ray can be externally detected for image reconstruction[6].

This work was supported in part by the National Natural Science Foundation of China (61201346). *Asterisk indicates corresponding author*

P. Feng is with the Key Laboratory of Optoelectronic Technology and Systems, Ministry of Education, Chongqing University, Chongqing 400044 China and now a visiting scholar of Biomedical Imaging Cluster, Department of Biomedical Engineering, Rensselaer Polytechnic Institute, Troy, NY 12180 USA (e-mail: coe-fp@cqu.edu.cn;).

*G. Wang and W.X. Cong are with Biomedical Imaging Cluster, Department of Biomedical Engineering, Rensselaer Polytechnic Institute, Troy, NY 12180 USA (e-mail: wangg6@rpi.edu; congw@rpi.edu).

B. Wei is with the Key Laboratory of Optoelectronic Technology and Systems, Ministry of Education, Chongqing University, Chongqing 400044 China (e-mail: weibiao@cqu.edu.cn).

Traditionally, XFCT requires monochromatic synchrotron x-rays to determine a spatial distribution of various elements[7]-[9], such as Gold(Au), Iodine (I) and Gadolinium (Gd). However, the synchrotron-based XFCT technique is unsuitable for *in vivo* studies in a typical laboratory setting. In 2010, Cheong et al. reported XFCT experiments with a small animal-sized object containing gold nanoparticles (GNPs) using a polychromatic X-ray source at 110 kVp[10]. Meanwhile, the photon-counting spectral detector which can separately record photons at different energies has been used in K-edge CT. With state-of-the-art Medipix-3 spectral detectors, the sensitivity and image quality of K-edge imaging can be improved with the specification of energy threshold setting[11][14]. Recently, Bazalova et al. reported a Monte Carlo XFCT study based on energy resolving detector and compared XFCT with K-edge CT, showing that XFCT outperforms to K-edge CT in terms of contrast-to-noise ratio (CNR) if the concentration of contrast agent is below 0.4%. In their work, the constraint of a zero attenuation background was assumed for incident and emitted fluorescence x-rays for use of the filtered backprojection (FBP) method to produce XFCT image [12].

In this paper, we combine analytic and numerical studies to compare XFCT and K-edge CT in terms of image resolution, noise, signal-to-noise ratio(SNR) and CNR. Two kinds of phantoms used in our work represent multiple concentration levels and feature sizes. XFCT and K-edge CT reconstructions are iteratively performed after attenuation correction. Data are synthesized assuming GNPs as the contrast agent at different concentrations. In comparison, K-edge CT[14]-[16] is also simulated in the identical experimental setting.

The outline of this paper is as follows: In Section II, we review the principle of XFCT, introduce K-edge CT and derive the SNR. In Section III, we compare XFCT and K-edge CT in terms of SNR. In Section IV, we describe numerical tests with phantoms containing multiple contrast concentration levels and feature sizes. In Section V, we present simulation results and discuss them qualitatively and quantitatively. Finally, in Section VI, we address relevant issues and conclude the paper.

## II. PHYSICAL MODELING

*A. XFCT*

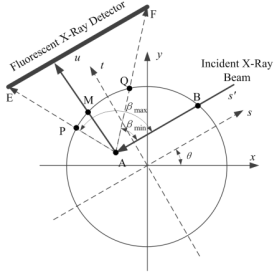

Fig. 1. Geometry of XFCT showing the object and detection coordinate systems respectively.

The physics and imaging model of XFCT are well known[8][9]. The geometry of XFCT is presented in Figure 1. While the $xy$-coordinate system is attached to an object, the $st$-coordinate system is spun with the data acquisition system, and can be at any instant obtained by rotating the $xy$-coordinate system by an angle $\theta$ counterclockwise. That is, the relationship between the two coordinate systems can be expressed as

$$s = x\cos\theta + y\sin\theta, \quad t = -x\sin\theta + y\cos\theta \quad (1)$$

In the $st$-coordinate system, let us consider the fluorescence contribution from a contrast agent at a position A on the primary beam line. This process can be divided into three steps: (1) the incident x-ray beam from position B arrives at position A; (2) the fluorescence x-rays are emitted uniformly from position A when incident x-ray beam interacts with contrast agent; and (3) a detector records x-ray fluorescence signals coming from position A.

Some fluorescence photons emitted from position A will travel through the phantom, and some photons will be absorbed. The intensity of the fluorescence x-rays measured by the detector can be formulated as,

$$I_{AEF}^{f} = I_0 e^{-\int_{-\infty}^{s} \mu^I(s',t)ds'} \omega\mu_{ph}\rho(s,t) \int_{\beta_{min}}^{\beta_{max}} e^{-\int_0^{\infty} \mu^F(s-u\cos\beta, s+u\sin\beta)du} d\beta \Delta s \quad (2)$$

where $I_0$ is the initial intensity of the incident x-ray beam or the mean number of incident x-ray photons, $\mu^I$ and $\mu^F$ are the linear attenuation coefficients for the incident and fluorescence x-rays respectively, $\rho(s,t)$ is the concentration of GNPs, $\mu_{ph}$ is the photoelectric mass absorption coefficient of GNPs, $\omega$ is the yield of fluorescence x-rays, $\beta$ is the solid angle, and $[\beta_{min}, \beta_{max}]$ is the solid angle range covered by the detector. The total intensity of the fluorescence x-ray signal reaching the detector for an incident x-ray beam is obtained by integrating $I_{AEF}^f$ along the primary x-ray direction as

$$I^f(\theta,t) = \omega\mu_{ph}I_0 \int_{s_{min}}^{s_{max}} [e^{-\int_{-\infty}^{s} \mu^I(s',t)ds'}] \rho(s,t)[\int_{\beta_{min}}^{\beta_{max}} e^{-\int_0^{\infty} \mu^F(s-u\cos\beta, s+u\sin\beta)du} d\beta]ds \quad (3)$$

Eq. (3) can be rewritten as

$$I^f(\theta,t) = \int_{s_{min}}^{s_{max}} f(\theta,s,t)g(\theta,s,t)\rho(s,t)ds \quad (4)$$

where

$$f(\theta,s,t) = I_0 e^{-\int_{-\infty}^{s} \mu^I(s',t)ds'} \quad (5)$$

$$g(\theta,s,t) = \omega\mu_{ph} \int_{\beta_{min}}^{\beta_{max}} e^{-\int_0^{\infty} \mu^F(s-u\cos\beta, s+u\sin\beta)du} d\beta \quad (6)$$

Note that $\theta$ and $t$ are variables denoting the angle and the offset of an incident x-ray respectively. For given $\theta$ and $t$ values, $f(\theta,s,t)$ reflects the attenuation process of the incident x-ray from position B to position A, and $g(\theta,s,t)$ is for the fluorescence emission from position A to the detector. With Eq. (4), the measurement $I^f(\theta,t)$ is directly linked to the contrast concentration distribution $\rho(s,t)$, assuming that $\mu^F(s,t)$ and $\mu^I(s,t)$ are both known. Hence, we can reconstruct $\rho(s,t)$ from $I^f(\theta,t)$.

It is well known that the number of photons emitted from an x-ray source is a random variable $I_{inc}$ which obeys a Poisson distribution[1]:

$$P\{I_{inc} = n\} = \frac{e^{-\bar{I}_{inc}}}{n!}(\bar{I}_{inc})^n \quad (7)$$

where $\bar{I}_{inc}$ is the mean value of random variable $I_{inc}$. From Beer-Lambert's law, the number of primary x-ray photons $I_A$ reaching to position A can be expressed as,

$$I_A = I_{inc} e^{-\int_{-\infty}^{s} \mu^I(s',t)ds'} \quad (8)$$

From Eq. (8), the number of x-ray photons $I_A$ arriving at position A also obeys a Poisson distribution[1],

$$P\{I_A = m\} = \frac{e^{-\bar{I}_A}}{m!}(\bar{I}_A)^m \quad (9)$$

Where $\bar{I}_A$ is the mean value of $I_A$. When x-ray photons interact with contrast agent at position A, the emitted x-ray fluorescence photons travel through the object. The number of survival fluorescence photons obeys the binomial distribution $B(k,m,p)$. Hence, the probability distribution of x-ray fluorescence photons $I_A^f$ recorded by detector can be written as follows,

$$P\{I_A^f = k\} = \sum_{l=k}^{\infty} P\{I_A = m\}B(k,m,p) = e^{-\bar{I}_A^f} \frac{(\bar{I}_A^f)^k}{k!} \quad (10)$$

Eq. (10) indicates that the probability distribution of $I_A^f$ is a Poisson distribution with a mean value $\bar{I}_A^f = \bar{I}_A * P$. Because the superposition of independent Poisson distribution is still a Poisson distribution, from Eq. (4) we obtain that the number of recorded x-ray fluorescence photons $\zeta$ obeys a Poisson distribution with expectation and variance:

$$E(\zeta) = \lambda^F \quad (11)$$

$$D(\zeta) = \lambda^F \quad (12)$$

where $\lambda^F$ is the mean value of $\zeta$. Hence, the signal-to-noise ratio (SNR) of x-ray fluorescence signal recorded by detector can be defined as:

$$SNR_F = \frac{E(\zeta)}{\sqrt{D(\zeta)}} = \sqrt{\lambda^F} \quad (13)$$

*B. K-edge CT*

K-edge CT depends on the two energy bins on both sides of a K-edge of a relatively high atomic number material. Let $\mu_{Au}(s,t,E)$ be the linear attenuation coefficient function of a GNPs solution at a position $(s,t)$ and energy $E$, and $\bar{\mu}_{Au}^R(s,t,W)$ be the average linear attenuation coefficient within an energy bin after the K-edge energy, and $\bar{\mu}_{Au}^{L}(s,t,W)$ the counterpart within another energy bin before the K-edge. $\mu_{Bk}(s,t,E)$, $\bar{\mu}_{Bk}^{R}(s,t,W)$ and $\bar{\mu}_{Bk}^{L}(s,t,W)$ are corresponding parameters for a background media. We have

$$\bar{\mu}_{Au}^{R}(s,t,W) = \frac{1}{W}\int_{K^{+}}^{K+w}\mu_{Au}(s,t,E)dE \quad (14)$$

$$\bar{\mu}_{Au}^{L}(s,t,W) = \frac{1}{W}\int_{K-w}^{K^{-}}\mu_{Au}(s,t,E)dE \quad (15)$$

$$\bar{\mu}_{Bk}^{R}(s,t,W) = \frac{1}{W}\int_{K^{+}}^{K+w}\mu_{Bk}(s,t,E)dE \quad (16)$$

$$\bar{\mu}_{Bk}^{L}(s,t,W) = \frac{1}{W}\int_{K-w}^{K^{-}}\mu_{Bk}(s,t,E)dE \quad (17)$$

where $W$ is the optimized width of the energy bin according reference[14] which shows that we can get the highest signal difference noise ratio(SDNR) for K-edge CT imaging, $K^{-}$ and $K^{+}$ are the left and right limits of the K-edge respectively.

Given a GNPs concentration, we have[16]

$$\mu_{Au}(s,t,E) = \rho(s,t)\beta_{Au}(E)Z_{Au} + [1-\rho(s,t)]\beta_{Bk}(E)Z_{Bk} \quad (18)$$

$$\mu_{Bk}(s,t,E) = \beta_{Bk}(E)Z_{Bk} \quad (19)$$

where $\rho(s,t)$ is the concentration of a GNPs solution, $\beta_{Au}(E)$ and $\beta_{Bk}(E)$ are the mass attenuation coefficients of Gold and background media[18] at energy level $E$, $Z_{Au}$ and $Z_{Bk}$ are the densities of gold and background media, respectively.

Based on the definition of K-edge imaging, we have

$$I_{Rs} = I_{inc}e^{-\int_{S_{min}}^{S_{max}}\bar{\mu}_{Au}^{R}(s,t,W)ds} \quad (20)$$

$$I_{Ls} = I_{inc}e^{-\int_{S_{min}}^{S_{max}}\bar{\mu}_{Au}^{L}(s,t,W)ds} \quad (21)$$

With Eqs. (20) and (21), we obtain

$$\ln\left(\frac{I_{Ls}}{I_{Rs}}\right) = \ln\left(\frac{I_{inc}e^{-\int_{S_{min}}^{S_{max}}\bar{\mu}_{Au}^{L}(s,t,w)ds}}{I_{inc}e^{-\int_{S_{min}}^{S_{max}}\bar{\mu}_{Au}^{R}(s,t,w)ds}}\right) \quad (22)$$

$$= \int_{S_{min}}^{S_{max}}\bar{\mu}_{Au}^{R}(s,t,W)ds - \int_{S_{min}}^{S_{max}}\bar{\mu}_{Au}^{L}(s,t,W)ds$$

According to the definition of the average linear attenuation coefficients, from Eqs. (14) and (15), Eq.(22) can be written as:

$$\ln\left(\frac{I_{Ls}}{I_{Rs}}\right) = \frac{1}{W}\int_{S_{min}}^{S_{max}}\int_{K^{+}}^{K+w}\mu_{Au}(s,t,E)dEds - \frac{1}{W}\int_{S_{min}}^{S_{max}}\int_{K-w}^{K^{-}}\mu_{Au}(s,t,E)dEds$$

Based on Eqs. (18) and (19), we have

$$\ln\left(\frac{I_{Ls}}{I_{Rs}}\right) = \frac{1}{W}\int_{S_{min}}^{S_{max}}\rho(s,t)Z_{Au}[\int_{K^{+}}^{K+w}\beta_{Au}(E)dE - \int_{K-w}^{K^{-}}\beta_{Au}(E)dE]ds$$

$$+ \frac{1}{W}\int_{S_{min}}^{S_{max}}[1-\rho(s,t)]Z_{Au}[\bar{\mu}_{Bk}^{R}(s,t,W) - \bar{\mu}_{Bk}^{L}(s,t,W)]ds$$

For a given energy bin, the average linear attenuation coefficients of background media are almost same on both sides of K-edge, i.e., $\bar{\mu}_{Bk}^{R}(s,t,W) \approx \bar{\mu}_{Bk}^{L}(s,t,W) = \bar{\mu}_{Bk}(s,t,W)$, we have

$$\ln\left(\frac{I_{Ls}}{I_{Rs}}\right) \approx \frac{1}{W}\int_{S_{min}}^{S_{max}}\rho(s,t)Z_{Au}[\int_{K^{+}}^{K+w}\beta_{Au}(E)dE - \int_{K-w}^{K^{-}}\beta_{Au}(E)dE]ds$$

$$= \left\{\frac{Z_{Au}}{W}[\int_{K^{+}}^{K+w}\beta_{Au}(E)dE - \int_{K-w}^{K^{-}}\beta_{Au}(E)dE]\right\}\int_{S_{min}}^{S_{max}}\rho(s,t)ds$$

Let $Co = \left\{\frac{Z_{Au}}{W}[\int_{K^{+}}^{K+w}\beta_{Au}(E)dE - \int_{K-w}^{K^{-}}\beta_{Au}(E)dE]\right\}$, we have

$$\ln\left(\frac{I_{Ls}}{I_{Rs}}\right) \approx Co\int_{S_{min}}^{S_{max}}\rho(s,t)ds$$

Thus, we get a relationship between the Radon transform of a GNPs concentration $\rho(s,t)$ and the logarithm of the ratio between measured photon intensities on both sides of the K-edge. Eq. (22) can be used to reconstruct a GNPs distribution from measured data of K-edge imaging.

## III. SNR ANALYSIS OF K-EDGE CT AND XFCT

From Section II, we obtain that the number of x-ray fluorescence photons hitting a detector obeys Poisson distribution, and the SNR of x-ray fluorescence imaging is $\sqrt{\lambda^{F}}$ where $\lambda^{F}$ is the average number of detected x-ray fluorescence photons. On the other hand, for x-ray K-edge CT, when incident x-ray beams pass through the object, the number of detected photons relies on the contrast agent concentration $\rho$ and its K-edge energy. The number of detected photons $\xi$ with an energy bin before the K-edge can be described by a Poisson distribution with mean value $\lambda$:

$$P\{\xi = n\} = e^{-\lambda}\frac{\lambda^{n}}{n!} \quad (23)$$

Similarly, the number of detected photons $\eta$ after the K-edge also obeys a Poisson distribution with mean value $\lambda_{R}$

$$P\{\eta = n\} = e^{-\lambda_{R}}\frac{(\lambda_{R})^{n}}{n!} \quad (24)$$

Let us define $x = \lambda_{R}/\lambda$, clearly, $0 < x < 1$. Based on Eq. (22) we can define the signal-to-noise ratio of x-ray K-edge CT($SNR_{K}$):

$$SNR_{K}(x) = \frac{E[\ln(\xi),\lambda] - E[\ln(\eta),x\lambda]}{\sqrt{D[\ln(\xi),\lambda]} + \sqrt{D[\ln(\eta),x\lambda]}} \quad (25)$$

where $\ln(\xi)$ obeys logarithm-Poisson distribution which is defined as,

$$\begin{cases} P\{\ln(\xi) = \ln(1)\} = e^{-\lambda}\frac{\lambda}{1!}, \\ P\{\ln(\xi) = \ln(n)\} = e^{-\lambda_{R}}\frac{\lambda_{R}^{n}}{n!}, \quad n = 2, 3, \cdots, n \in Z^{+} \end{cases} \quad (26)$$

$E[\ln(\xi),\lambda]$ and $D[\ln(\xi),\lambda]$ are its expectation and variance, respectively. Accordingly, $\ln(\eta)$ is a random variable and its distribution is similar to $\ln(\xi)$; $E[\ln(\eta),x\lambda]$ and $D[\ln(\eta),x\lambda]$ are the expectation and variance of $\ln(\eta)$, respectively.

Obviously, the definition of $SNR_{K}(x)$ relies on the concentration of contrast agent. When $x$ approaches 1, GNPs become rather sparse in the object, and $SNR_{K}(x)$ approaches 0. This is consistent to intuition – no GNP no SNR. When $x$ approaches 0, GNPs with high concentration attenuate most of x-ray photons in the phantom. In this case, the detected x-ray photons before the K-edge of GNPs represent the measured signal, although few photons can be detected after the K-edge of GNPs. In other words, the fact that no photon is seen after the K-edge helps confirm the information contained in the measurement made before the K-edge. Furthermore, the function $\text{SNR}_K(x)$ has following two properties:

**Lemma 1**: $\text{SNR}_K(x)$ is a monotonically decreasing function.

Proof: See Appendix A.

**Lemma 2**: $\text{SNR}_K(0) > \sqrt{\lambda}$.

Proof: See Appendix B.

Combining Lemmas 1 and 2, we conclude that there is a threshold for the GNPs concentration in the context of XFCT and K-edge CT. When the GNPs concentration is less than the threshold, XFCT has a better SNR than K-edge CT; vice versa. As a matter of fact, by Lemmas 1 and 2, $\text{SNR}_K(x)$ is monotonically decreasing, $\text{SNR}_K(0) > \sqrt{\lambda}$ and $\text{SNR}_K(1) = 0$. In this case, we assume that incident x-ray photons are same for both imaging modalities at every projection view. It's obvious that the energy of x-ray fluorescence excited by primary x-ray is same to the absorbed energy of contrast element at its K-edge which indicates the detected x-ray fluorescence photons should be smaller than transmitted x-ray photons, i.e. $\lambda^F < \lambda$. So, there must be one and only one point $x_{Th}$ in the interval (0,1) such that $\text{SNR}_K(x_{Th}) = \sqrt{\lambda^F}$ by the intermediate value theorem. At and only at this threshold value $x_{Th}$, $\text{SNR}_K$ is the same as $\text{SNR}_F$. For $0 < x < x_{Th}$, $\text{SNR}_K > \text{SNR}_F$ and K-edge CT is superior to XFCT. For $x_{Th} < x < 1$, $\text{SNR}_K < \text{SNR}_F$, and XFCT outperforms K-edge CT in terms of SNR.

## IV. NUMERICAL SIMULATION

### A. Phantoms

The numerical simulation geometry of XFCT is shown in Fig 2. The incident x-ray flux is $10^6$ photons/mm$^2$, with a pencil beam cross section of 0.25 mm$^2$. The detector whose length is equal to the diameter of phantom is placed parallel to the direction of incident x-ray beam, the distance between detector and the border of phantom is 10 millimeter. The phantoms are scanned with the first generation CT manner and the phantom's translational step is 0.5mm and its rotational step is 3.6°. We have two kinds of phantoms: Multi-concentration (Fig 2(a)) for contrast resolution and Multi-size (Fig 2(b)) for spatial resolution. As shown in Fig 2(a), multi-concentration phantoms has 3 phantoms (Phantom 1, Phantom 2 and Phantom 3), each phantom is a circular object of 64 millimeter diameter which is discretized into a 128×128 matrix, i.e., the pixel size is 0.25 mm$^2$. It contains 7 disks (A to G) of 10 mm diameter and the former 6 disks(A to F) are placed uniformly along a circle whose diameter is 45mm. The 7$^{th}$ disk(G) is placed in the center of the phantom. For the multi-concentration phantoms, 7 disks which are filled with contrast agent of 7 different concentrations are contrast agent test ROIs and background is full of water. Here we use GNPs solution as the contrast agent, The densities of gold and water are 19.34g/cm$^3$ and 1g/cm$^3$, respectively. Table I lists all the concentrations of contrast agent from 0.1% to 2.1%.

As shown in Fig 2(b), the multi-size phantom is designed to evaluate the spatial-resolving performance of XFCT. It also has

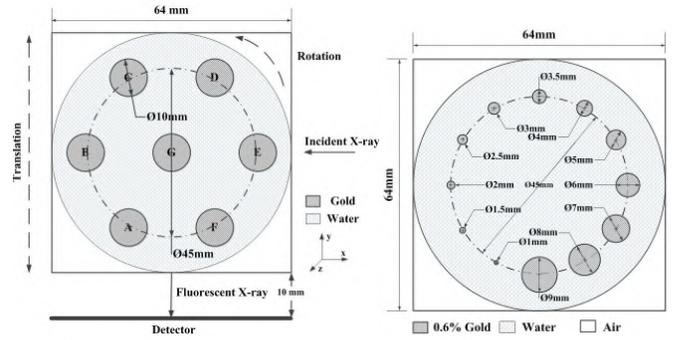

(a) Phantom 1 to 3: Multi-concentration   (b)Phantom 4:Multi-size

Fig. 2. Simulation setup of x-ray fluorescence tomography imaging

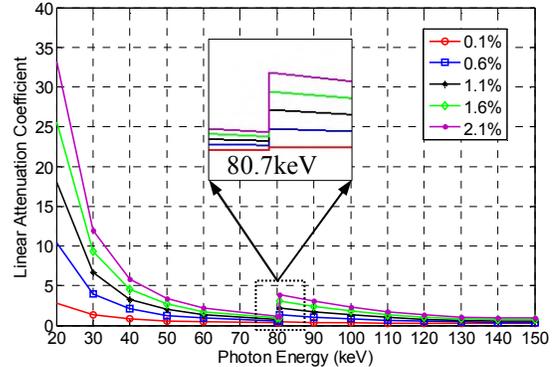

Fig. 3. Linear attenuation coefficient curve with different concentrations

TABLE I
THE CONCENTRATIONS OF GNPS AND
CORRESPONDING LINEAR ATTENUATION COEFFICIENTS

| Pha. No. | Disk | | A | B | C | D | E | F | G |
|---|---|---|---|---|---|---|---|---|---|
| 1 | $\rho$(%) | | 0.1 | 0.4 | 0.7 | 1.0 | 1.3 | 1.6 | 1.9 |
|  | $\mu$ | L | 0.331 | 0.450 | 0.569 | 0.688 | 0.807 | 0.926 | 1.045 |
|  |  | R | 0.446 | 0.942 | 1.437 | 1.933 | 2.429 | 2.924 | 3.419 |
| 2 | $\rho$(%) | | 0.2 | 0.5 | 0.8 | 1.1 | 1.4 | 1.7 | 2.0 |
|  | $\mu$ | L | 0.371 | 0.489 | 0.609 | 0.728 | 0.847 | 0.966 | 1.084 |
|  |  | R | 0.611 | 1.107 | 1.603 | 2.098 | 2.594 | 3.089 | 3.585 |
| 3 | $\rho$(%) | | 0.3 | 0.6 | 0.9 | 1.2 | 1.5 | 1.8 | 2.1 |
|  | $\mu$ | L | 0.4104 | 0.529 | 0.648 | 0.767 | 0.886 | 1.005 | 1.124 |
|  |  | R | 0.776 | 1.272 | 1.768 | 2.263 | 2.758 | 3.255 | 3.750 |

Here Pha. No. is phantomn; $\rho$ is the concentration of contrast agent; $\mu$ is linear attenuation coefficient; L means "Left side" and R means "Right side";

a circular object of 64 mm diameter(128×128) and contains 12 small disks with different diameters from 1mm to 9mm. We use GNPs solution as the contrast agent too and all the small disks are filled with contrast agent of same concentration 0.6%.

According to the x-ray attenuation databases reported by the National Institute of Standards and Technology (NIST) [18], we can easily get all the linear attenuation coefficients curves $\mu(E)$ with different concentrations of GNPs. A few examples are plotted in Fig 3. For the K-edge tomographic imaging, we need to reconstruct GNPs distribution in the object based on Eq. (22). According to reference[14] and Eqs. (14) to (17), we set the energy bin $W$ equal to 3 keV for all concentration, the integration window is from 80.7 keV (K-edge energy of gold) to 83.7 keV for the right side and 77.7 keV to 80.7 for the left side. All the linear attenuation coefficients for different concentrations are also listed in Table I where "L" indicates integration window is before the K-edge energy and "R" is after the K-edge. $\rho$ is the concentration of contrast agent; $\mu$ is linear attenuation coefficient.

*B. Reconstruction Method*

Various methods can be used for XFCT and K-edge CT reconstruction: analytical reconstruction approach like filter backprojection (FBP), or the iterative reconstruction like algebraic reconstruction technique (ART) and simultaneous ART (SART)[1][19][20]; In this paper, the maximum likelihood method with expectation maximization (MLEM) is applied for reconstruction, the known formula in reference[21] can be written as follows:

$$\varphi_j^{(n+1)} = \frac{\varphi_j^{(n)}}{\sum_{i=1}^{M} h_{i,j}} \sum_{i=1}^{M} \frac{h_{i,j} p_i}{\sum_{k=1}^{N} h_{i,k} \varphi_k^{(n)}}, \ 1 \leq j \leq N, 1 \leq n \leq T \quad (27)$$

Where $h_{i,j} (1 \leq i \leq M, 1 \leq j \leq M, i \in \mathbf{Z}^+)$ is discretized projection matrix $H$ which is defined as

$$H = H(\theta, s, t) = f(\theta, s, t) g(\theta, s, t) \quad (28)$$

$M$ is the total projection number and $N$ is the total pixel number of a phantom that will be reconstructed. $p_i$ is the detected fluorescence x-ray count for the *i*th projection. $\varphi_j^{(n+1)}$ is the *j*th reconstructed pixel value at the (*n*+1)th iteration. So *i* determines the index of projection, *j* the pixel index of the reconstructed distribution of concentration. In this case, $M$ equal to $128 \times (180/3.6) = 6400$ and $N$ is $128 \times 128 = 16384$. For total iteration number $T$, we set $T=200$, there is not specific formula about how to choose the total iteration number $T$, one can choose a number according the error convergence rate. For more detail, please refer to[18];

V. RESULTS

For multi-concentration phantom, we performed 10 groups of simulation for XFCT and K-edge CT, each group has 3 phantoms. In order to evaluate the reconstruction quality due to the influence of different imaging x-ray dose, we add 3 kinds of Gaussian noise to the detected fluorescence x-ray signal. All 3 noises are zero-mean and variance $\sigma^2$ are 20, 30 and 40, respectively. One example of all reconstructed XFCT and K-edge CT images for various noise variances are shown in Fig 4. It's evident that both in XFCT and K-edge CT, the quality of reconstructed images are becoming worse with the increase of noise variance. Especially when $\sigma^2=40$, the reconstructed images are interfered by the noise severely: non-uniform ROI and background, blurred edge of each disk, partially absence of phantom and some artifacts in the area of air. Furthermore, as shown in Fig5 (a) to (c) and (j) to (l), the disk with the lowest concentration contrast agent (disk A) in each phantom are hardly visible which means if the x-ray imaging dose is too low, the contrast agent with concentration from 0.1% to 0.3% are undetectable.

It's also obviously the quality of reconstructed images improve with the decrease of noise variance, especially for the uniformity of ROI and background. However, even the noise variance is 20, disk with 0.1% concentration contrast agent is also undetectable in XFCT and K-edge CT. When $\sigma^2=30$, for XFCT and K-edge CT, the lowest concentration contrast agent could be detected is 0.3%(Fig 4 (e)) and 0.4% (Fig 4 (m)), respectively. It shows that XFCT is more sensitive to the low concentration than K-edge CT partially due to background interfere of CT imaging and the subtraction operation of K-edge imaging which will augment variance of background noise.

Fig 4 shows in XFCT imaging the average reconstructed concentration value and variance of ROI as a function of noise variance and contrast agent concentration. It coincides with Fig 4 that variance of reconstructed value becomes smaller when noise variance decreases, which means the reconstructed value is more stable, uniform and accurate. Another thing we should notice is that no matter what the noise variance is, the variance of reconstructed concentration value increase with the increase of concentration. This determines that the reconstructed value for larger contrast concentration has bigger fluctuation compared with smaller concentration.

The reconstructed images are also evaluated as contrast-to-noise ratio (CNR) by calculating the ratio of the difference between the mean value of each ROI and background(Water) and square root of sum of variance of ROI and background. CNR is defined as

$$\mathrm{CNR} = \frac{\bar{\Psi}_{ROI} - \bar{\Psi}_{BK}}{\sqrt{\Phi_{ROI}^2 + \Phi_{BK}^2}} \quad (29)$$

Where $\bar{\Psi}_{ROI}$ and $\bar{\Psi}_{BK}$ are mean reconstructed values of ROI and background, $\Phi_{ROI}^2$ and $\Phi_{BK}^2$ are corresponding variances of ROI and background. The CNR as a function of contrast agent concentration and noise variance has been plotted in Fig 6. It's easy to find there is an approximately linear relationship between the CNR and contrast agent concentration for all noise variance conditions, and CNR increases much faster when concentration is smaller than 0.4% than it does when concentration is bigger than 0.4%, which coincides with what we mentioned above that XFCT is very sensitive to the change of low concentration. When $\sigma^2=20$ all CNRs are lower than 0.7 for concentration from 0.1% to 0.3%, but when $\sigma^2=30$, CNRs are lower than 0.7 for concentration only from 0.1% to 0.2%. So we can get a direct inference that 0.7 is the threshold value of CNR that determine whether this contrast agent is detectable.

For comparison, K-edge CT imaging is also performed. The average CNR difference between XFCT and K-edge CT as a function of noise variance and contrast concentration are shown in Fig 7. Just like results of reference[13], in general condition, for most of contrast agent concentration from 0.8% to 1.8%, there is a trend that CNR of reconstructed XFCT images is lower than CNR of K-edge CT images, and the difference increases with the increase of contrast agent concentrations. But we find there is a sudden increase when concentration is larger than 1.8%, we infer that for disks placed in the center of phantom(concentration from 1.9% to 2.1%), the position of which will influence the reconstructed value of XFCT and make CNR larger than it expected if contrast agent is not placed in center.

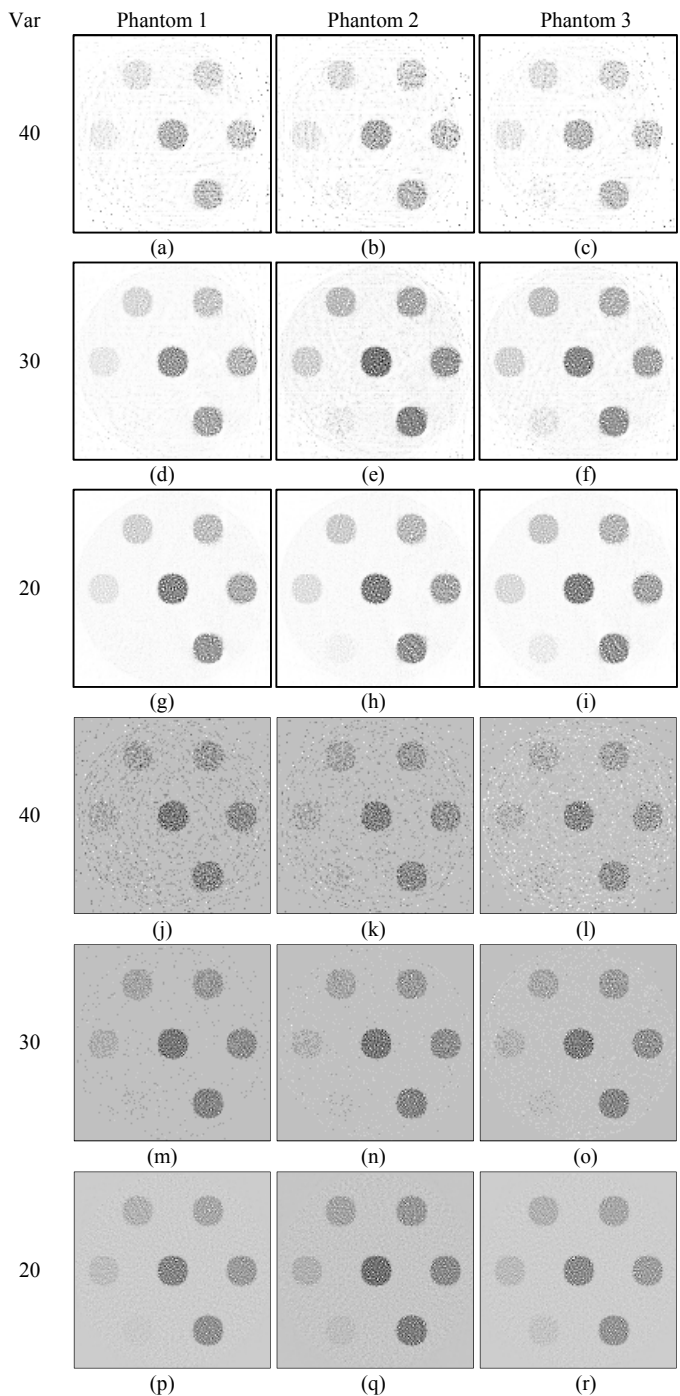

Fig.4. Reconstructed images of 3 parts with different noise variance. Top(a-i) is XFCT and bottom(j-r) is K-edge CT. All images are normalized for display

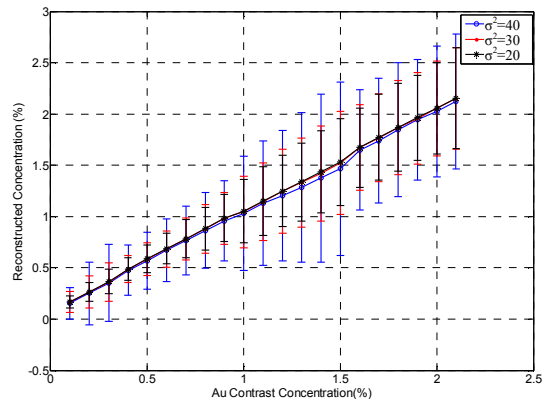

Fig 5 Average reconstructed concentration value as a function of noise variance and contrast concentration

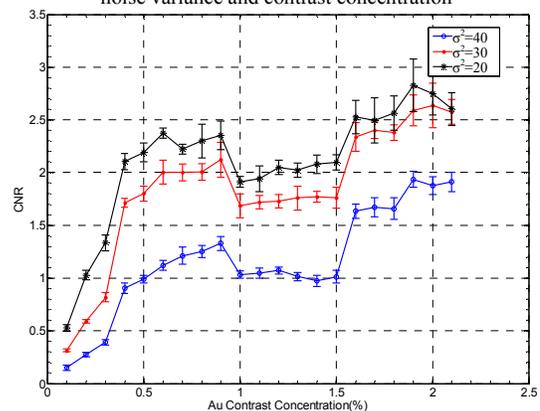

Fig 6 CNR for reconstructed XFCT image as a function of noise variance and contrast agent concentration

For Multi-size phantom, GNPs solution of 0.6% concentration as the contrast agent is filled in 12 small disks whose diameters are from 1 mm to 9 mm and water is as background. We reconstruct the images for XFCT and K-edge CT with noise $\sigma^2=30$, one example of reconstructed images is shown in Fig 8 and the average difference of CNR between XFCT and K-edge CT is also plotted in Fig 9. As shown in Fig 8, obviously XFCT can get better reconstructed images for smaller-size disks(diameters from 1mm to 3mm) compared with K-edge CT in terms of uniformity, edge, contrast. Due to the influence of noise, the difference of CNR has some fluctuation. But the fitted curve shows a trend that the difference between XFCT and K-edge becomes smaller and smaller with the increase of disk size. In other words, K-edge CT can get better results if size of ROI is bigger.

## VI. DISCUSSIONS AND CONCLUSION

This is a theoretical analysis and preliminary simulation for XFCT where many factors, such as the energy spectrum distribution of x-ray source, attenuation correction of phantom, other physical reaction except absorption and emission and detector efficiency, et al., can affect the final results. Nevertheless, our derivation and simulation is rigorous and can be applied to a class of phantoms.

Theoretically, we analyze the performance of XFCT and K-edge CT in terms of SNR under the circumstance that only Poisson noise is taken into consideration. The relationship of SNR and concentration of contrast agent is derived which shows that XFCT outperforms K-edge CT when concentration is lower than a threshold value and K-edge CT is superior to XFCT when concentration is higher than this value. Similar results of numerical simulation are also achieved. For contrast agent concentrations of GNPs from 0.8%–2.1%, we show that there is a trend K-edge CT outperforms XFCT in terms of CNR and the difference of CNR increase with the increase of contrast agent concentration. For concentrations from 0.5%~0.7%, it's a transitional area where CNR of both have small difference if we set ±0.1 as the threshold. And we also notice that transitional area will change with the noise variance, which is shown in Fig 7(a) as two small squares(solid and dashed). For concentrations from 0.1%~0.4%, CNR of XFCT images is higher than CNR of K-edge CT images. This can be attributed to the combination of higher noise in K-edge CT images compared to XFCT images and the imperfect subtraction of the background material. Also a fixed width of energy bin for K-edge CT imaging which does not automatically adjust with the contrast agent concentrations may not get the best reconstructed image. Unlike conventional transmission CT imaging, XFCT and K-edge imaging result in contrast-only images. CNR of XFCT does not depend on the background tissue type. On the other hand, the background tissue subtraction is not perfect in K-edge CT images, which causes a decrease in CNR for low concentrations compared to XFCT images.

Although the scattered fluorescence x-ray has already been introduced into our simulation with simple 1-order model, it does not change reconstructed image much due to the uniform distribution of contrast concentration. Another thing we are not very clear is that CNR has a sudden decrease from 0.9% to 1.0% and increases slowly between 1.0% and 1.5% which is shown in Fig 6, so more complicated phantoms and simulations will be investigated in the follow up study.

In conclusion, a theoretical and numerical comparison for XFCT and K-edge CT in terms of image resolution, noise, SNR and CNR has been proposed. From analytic study we can get that there is a threshold value in GNPs concentration and XFCT has a better SNR than K-edge CT if GNPs concentration is less than this value. Numerical simulations with multiple concentration level and feature sizes are also performed and experimental results show that XFCT is superior to K-edge CT when contrast concentration is lower than 0.4% which coincides with the theoretical analysis. Further step should be focused on taking more parameters into consideration to improve the accuracy.

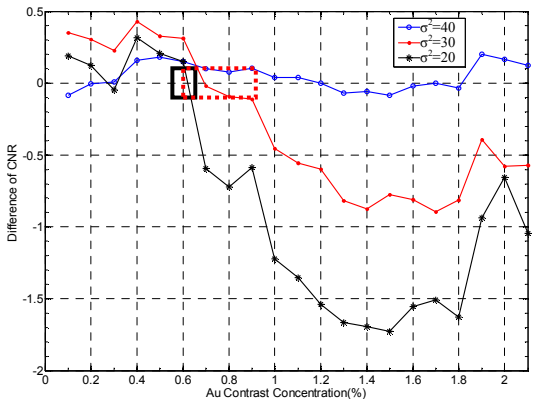

Fig 7 Difference of CNR(XFCT minus K-edge CT) as a function of concentration and noise variance, dashed square means transitional area for $\sigma^2=30$, solid square is for $\sigma^2=20$

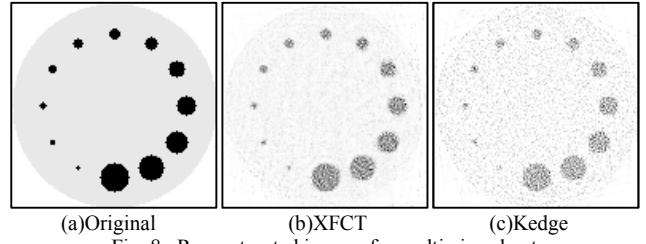

(a)Original     (b)XFCT     (c)Kedge
Fig. 8. Reconstructed images for multi-size phantom

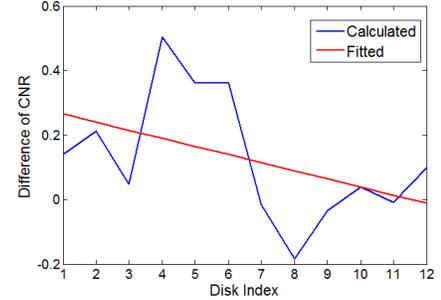

Fig. 9. Difference of CNR between XFCT and K-edge CT for multi-size phantom. X-coordinate is index of disk number. "1" corresponds disk with 1mm diameter, "12" corresponds disk with 9mm diameter.

APPENDIX

A. *Lemma 1:* $SNR_K(x)$ *is a monotonically decreasing function*

Proof: Let $SNR_K(x) = \Theta(x)$, then

$$\frac{d[\Theta(x)]}{dx} = \Theta'(x) = \frac{\tau_1 * \tau_2 - \tau_3 * \tau_4}{\tau_5} \quad (A.1)$$

where

$$\tau_1 = -2\left[\sqrt{D[\ln(\xi),\lambda]}\sqrt{D[\ln(\eta),\lambda x]} + D[\ln(\eta),\lambda x]\right]$$

$$\tau_2 = \frac{d}{dx}E[\ln(\eta),\lambda x]$$

$$\tau_3 = E[\ln(\xi),\lambda] - E[\ln(\eta),\lambda x]$$

$$\tau_4 = \frac{d}{dx}D[\ln(\eta),\lambda x]$$

$$\tau_5 = 2\sqrt{D[\ln(\xi),\lambda]}\left(\sqrt{D[\ln(\xi),\lambda]} + \sqrt{D[\ln(\eta),\lambda x]}\right)^2$$

Furthermore,

$$\tau_2 = \frac{d}{dx}E[\ln(\eta),\lambda x] = \sum_{n=1}^{\infty}\ln(1+\frac{1}{n})e^{-(\lambda x)}\frac{(\lambda x)^n}{n!} \quad (A.2)$$

$$\tau_4 = \frac{d}{dx}D[\ln(\eta),\lambda x]$$

$$= D[\ln(\eta+1),\lambda x] - D[\ln(\eta),\lambda x] + \left[\sum_{n=1}^{\infty}\ln(1+\frac{1}{n})e^{-(\lambda x)}\frac{(\lambda x)^n}{n!}\right]^2 \quad (A.3)$$

By the differential Mean Value Theorem, we have

$$\tau_3 = E[\ln(\xi),\lambda] - E[\ln(\eta),\lambda x]$$

$$= (1-r)\sum_{n=1}^{\infty}\ln(1+\frac{1}{n})e^{-(\lambda r)}\frac{(\lambda r)^n}{n!} \quad ,x<r<1,r\in R^+ \quad (A.4)$$

It can be easily shown that $\tau_1 < 0$, $\tau_2 > 0$, $\tau_3 > 0$ and $\tau_5 > 0$

If $\tau_4 \geq 0$, $\tau_1 * \tau_2 - \tau_3 * \tau_4$ should be smaller than 0, i.e.,

$$\Theta'(x) = \frac{\tau_1 * \tau_2 - \tau_3 * \tau_4}{\tau_5} < 0$$

If $\tau_4 < 0$, from Eq. (A.3), we have

$$-\tau_4 = |\tau_4|$$
$$= D[\ln(\eta), \lambda x]$$
$$- D[\ln(\eta+1), \lambda x] - \left[\sum_{n=1}^{\infty}\ln(1+\frac{1}{n})e^{-(\lambda x)}\frac{(\lambda x)^n}{n!}\right]^2 \quad (A.5)$$
$$< D[\ln(\eta), \lambda x] < D[\ln(\eta), \lambda x] + \sqrt{D[\ln(\xi), \lambda]D[\ln(\eta), \lambda x]}$$

The right side of the above inequality is $-\tau_1/2$, we can derive that

$$0 < -\tau_4 < -\tau_1/2$$

Then, $\tau_1 < 2\tau_4 < \tau_4 < 0$, i.e. $\tau_1 < \tau_4 < 0$. because function $\ln(1+1/n)$ in Eqs. (A.2) and (A.4) is a decrease function with respect to $n$, and $0 < x < r < 1$, we have $\tau_2 > \tau_3 > 0$. Hence, we obtain two following inequalities:

$$\tau_1 * \tau_3 < \tau_3 * \tau_4 < 0$$
$$\tau_1 * \tau_2 < \tau_1 * \tau_3 < 0$$

That is, $\tau_1 * \tau_2 < \tau_3 * \tau_4 < 0$, which means that

$$\Theta'(x) = \frac{\tau_1 * \tau_2 - \tau_3 * \tau_4}{\tau_5} < 0$$

In the following, let us numerically verify Lemma 1. First, the parameters were set up in consistence to real experiments. Table A.I shows the results of $\Theta'(x)$ for different $x$ and $\lambda$. For common micro-focus polychromatic x-ray sources, the average x-ray photon count rate ($\lambda$) is approximate $10^6$ counts/mm²/s and often decreased by 10~100 times if a filter is applied to the incident beam[11-12]. Hence, 4 different $\lambda$ values were tested, including $1\times10^3$, $1\times10^4$, $1\times10^5$ and $1\times10^6$ counts/mm²/s. It can be seen in Table A.I that $\Theta'(x) < 0$ for $0 < x \leq 1$. In other words, $\Theta(x)$ is monotonically decreasing over the interval [0, 1], and $\Theta(x)|_{x\in[0,1], x\neq 0}$ should be smaller than $\Theta(0)$.

TABLE A.I
NUMERICAL RESULTS OF $\Theta'(x)$ FOR DIFFERENT $x$ AND $\lambda$

| | | $\lambda$(cnt/mm²/s) | | | |
|---|---|---|---|---|---|
| $\Theta'(x)$ | | $1\times10^6$ | $1\times10^5$ | $1\times10^4$ | $1\times10^3$ |
| | 0 | 0 | 0 | 0 | 0 |
| | 0.1 | -294.47 | -95.10 | -29.82 | -8.58 |
| | 0.2 | -721.33 | -216.93 | -68.53 | -21.49 |
| | 0.3 | -679.59 | -227.91 | -72.06 | -22.72 |
| | 0.4 | -787.97 | -220.38 | -69.66 | -22.00 |
| $x$ | 0.5 | -670.37 | -208.69 | -66.02 | -20.86 |
| | 0.6 | -552.58 | -196.90 | -62.27 | -19.68 |
| | 0.7 | -672.59 | -185.78 | -58.76 | -18.57 |
| | 0.8 | -547.79 | -175.58 | -55.54 | -17.56 |
| | 0.9 | -523.57 | -166.43 | -52.63 | -16.64 |
| | 1.0 | -500.00 | -158.11 | -50.00 | -15.81 |

B. *Lemma 2*: $\Theta(0) > \sqrt{\lambda}$.

Proof: According to Eq. (A.1), When $x$ approaches 0 which means no photons can emitted from the phantom, i.e. the probability of 0 x-ray photon is 1, it's easy to derive $E[\ln(\eta), x\lambda]=0$ and $D[\ln(\eta), x\lambda]=0$, the SNR $\Theta(x)$ is reduced to

$$\Theta(0) = \frac{E[\ln(\xi), \lambda]}{\sqrt{D[\ln(\xi), \lambda]}} = \frac{\sum_{n=1}^{\infty}\ln(n)e^{-\lambda}\frac{\lambda^n}{n!}}{\sqrt{\sum_{n=1}^{\infty}\ln^2(n)e^{-\lambda}\frac{\lambda^n}{n!} - [\sum_{n=1}^{\infty}\ln(n)e^{-\lambda}\frac{\lambda^n}{n!}]^2}} \quad (B.1)$$

It can be shown that

$$\sum_{n=1}^{\infty}\ln^2(n)e^{-\lambda}\frac{\lambda^n}{n!} = \lambda\sum_{n=1}^{\infty}\ln(n)\frac{\ln(n)}{n}e^{-\lambda}\frac{\lambda^{n-1}}{(n-1)!} \quad (B.2)$$

According to Lemma 2.1 in Reference[23] that if $X$ is a random variable, $T$ and $R$ are continuous functions, and if $T$ is monotonically increasing and $R$ monotonically decreasing, then

$$E[T(X)R(X)] \leq E[T(X)]E[R(X)] \quad (B.3)$$

In our case, $n$ is random variable, $T(n)=\ln(n+1)$, $R(n)=\ln(n+1)/(n+1)$, we have

$$\sum_{n=1}^{\infty}\ln^2(n)e^{-\lambda}\frac{\lambda^n}{n!} = \lambda\sum_{n=0}^{\infty}\ln(n+1)\frac{\ln(n+1)}{n+1}e^{-\lambda}\frac{\lambda^n}{n!}$$
$$\leq \lambda\sum_{n=0}^{\infty}\ln(n+1)e^{-\lambda}\frac{\lambda^n}{n!}\sum_{n=0}^{\infty}\frac{\ln(n+1)}{n+1}e^{-\lambda}\frac{\lambda^n}{n!}$$
$$= \sum_{n=0}^{\infty}\ln(n+1)e^{-\lambda}\frac{\lambda^n}{n!}\sum_{n=1}^{\infty}\ln(n+1)e^{-\lambda}\frac{\lambda^{n+1}}{(n+1)!}$$

Since $\ln(n+1) = \ln(\frac{n+1}{n}n) = \ln(\frac{1}{n}+1) + \ln(n)$, we have

$$\sum_{n=1}^{\infty}\ln^2(n)e^{-\lambda}\frac{\lambda^n}{n!} \leq \sum_{n=1}^{\infty}[\ln(\frac{1}{n}+1) + \ln(n)]e^{-\lambda}\frac{\lambda^n}{n!}\sum_{n=1}^{\infty}\ln(n)e^{-\lambda}\frac{\lambda^n}{n!}$$
$$= \sum_{n=1}^{\infty}\ln(\frac{1}{n}+1)e^{-\lambda}\frac{\lambda^n}{n!}\sum_{n=1}^{\infty}\ln(n)e^{-\lambda}\frac{\lambda^n}{n!} + [\sum_{n=1}^{\infty}\ln(n)e^{-\lambda}\frac{\lambda^n}{n!}]^2$$

Hence,

$$\sum_{n=1}^{\infty}\ln^2(n)e^{-\lambda}\frac{\lambda^n}{n!} - \left[\sum_{n=1}^{\infty}\ln(n)e^{-\lambda}\frac{\lambda^n}{n!}\right]^2$$
$$\leq \left[\sum_{n=1}^{\infty}\ln(n)e^{-\lambda}\frac{\lambda^n}{n!}\right]\left[\sum_{n=1}^{\infty}\ln(\frac{1}{n}+1)e^{-\lambda}\frac{\lambda^n}{n!}\right] \quad (B.4)$$

Denoting

$$Q(\lambda) = (1+\lambda)\left[\sum_{n=1}^{\infty}\ln(n)e^{-\lambda}\frac{\lambda^n}{n!}\right]^2 - \lambda\left[\sum_{n=1}^{\infty}\ln^2(n)e^{-\lambda}\frac{\lambda^n}{n!}\right] \quad (B.5)$$

We have

$$\frac{dQ}{d\lambda} = 2(1+\lambda)\left[\sum_{n=1}^{\infty}\ln(n)e^{-\lambda}\frac{\lambda^n}{n!}\right]\left[\sum_{n=1}^{\infty}\ln(n)\left(e^{-\lambda}\frac{\lambda^{n-1}}{(n-1)!} - e^{-\lambda}\frac{\lambda^n}{n!}\right)\right]$$
$$- \sum_{n=1}^{\infty}\ln^2(n)e^{-\lambda}\frac{\lambda^n}{n!} - \lambda\sum_{n=1}^{\infty}\ln^2(n)\left[e^{-\lambda}\frac{\lambda^{n-1}}{(n-1)!} - e^{-\lambda}\frac{\lambda^n}{n!}\right]$$
$$+ \left[\sum_{n=1}^{\infty}\ln(n)e^{-\lambda}\frac{\lambda^n}{n!}\right]^2$$

Furthermore

$$\frac{dQ}{d\lambda} = 2(1+\lambda)\left[\sum_{n=1}^{\infty}\ln(n)e^{-\lambda}\frac{\lambda^n}{n!}\right]\left[\sum_{n=1}^{\infty}\ln\left(1+\frac{1}{n}\right)e^{-\lambda}\frac{\lambda^n}{n!}\right]$$
$$- \lambda\sum_{n=1}^{\infty}\left[\ln^2(n+1) - \ln^2(n)\right]e^{-\lambda}\frac{\lambda^n}{n!} + \left[\sum_{n=1}^{\infty}\ln(n)e^{-\lambda}\frac{\lambda^n}{n!}\right]^2 \quad (B.6)$$
$$- \sum_{n=1}^{\infty}\ln^2(n)e^{-\lambda}\frac{\lambda^n}{n!}$$

The second item of above equality can be written as:

$$\lambda \sum_{n=1}^{\infty} \left[ \ln^2(n+1) - \ln^2(n) \right] e^{-\lambda} \frac{\lambda^n}{n!}$$

$$= \lambda \sum_{n=1}^{\infty} \ln(1+\frac{1}{n})\left[ \ln(n+1) + \ln(n) \right] e^{-\lambda} \frac{\lambda^n}{n!}$$

$$= \lambda \sum_{n=1}^{\infty} \ln(n+1)\ln(1+\frac{1}{n}) e^{-\lambda} \frac{\lambda^n}{n!} - \lambda \sum_{n=1}^{\infty} \ln(n)\ln(1+\frac{1}{n}) e^{-\lambda} \frac{\lambda^n}{n!}$$

$$+ 2\lambda \sum_{n=1}^{\infty} \ln(n)\ln(1+\frac{1}{n}) e^{-\lambda} \frac{\lambda^n}{n!}$$

$$= \lambda \sum_{n=1}^{\infty} \ln^2(1+\frac{1}{n}) e^{-\lambda} \frac{\lambda^n}{n!} + 2\lambda \sum_{n=1}^{\infty} \ln(n)\ln(1+\frac{1}{n}) e^{-\lambda} \frac{\lambda^n}{n!}$$

Then

$$\frac{dQ}{d\lambda}$$
$$= \left[ \sum_{n=1}^{\infty} \ln(n) e^{-\lambda} \frac{\lambda^n}{n!} \right]\left[ \sum_{n=1}^{\infty} \ln\left(1+\frac{1}{n}\right) e^{-\lambda} \frac{\lambda^n}{n!} \right] - \lambda \left[ \sum_{n=1}^{\infty} \ln^2(1+\frac{1}{n}) e^{-\lambda} \frac{\lambda^n}{n!} \right]$$
$$+ 2\lambda \left[ \sum_{n=1}^{\infty} \ln(n) e^{-\lambda} \frac{\lambda^n}{n!} \right]\left[ \sum_{n=1}^{\infty} \ln\left(1+\frac{1}{n}\right) e^{-\lambda} \frac{\lambda^n}{n!} \right]$$
$$- 2\lambda \sum_{n=1}^{\infty} \ln(n)\ln(1+\frac{1}{n}) e^{-\lambda} \frac{\lambda^n}{n!}$$
$$+ \left[ \sum_{n=1}^{\infty} \ln(n) e^{-\lambda} \frac{\lambda^n}{n!} \right]\left[ \sum_{n=1}^{\infty} \ln\left(1+\frac{1}{n}\right) e^{-\lambda} \frac{\lambda^n}{n!} \right] + \left[ \sum_{n=1}^{\infty} \ln(n) e^{-\lambda} \frac{\lambda^n}{n!} \right]^2$$
$$- \sum_{n=1}^{\infty} \ln^2(n) e^{-\lambda} \frac{\lambda^n}{n!}$$

From Eq. (B.3), we have

$$\frac{dQ}{d\lambda} \geq \gamma_1 + \gamma_2 + \gamma_3 \qquad (B.7)$$

where

$$\gamma_1 = \left[ \sum_{n=1}^{\infty} \ln(n)\ln\left(1+\frac{1}{n}\right) e^{-\lambda} \frac{\lambda^n}{n!} \right] - \lambda \left[ \sum_{n=1}^{\infty} \ln^2\left(1+\frac{1}{n}\right) e^{-\lambda} \frac{\lambda^n}{n!} \right] \quad (B.8)$$

$$\gamma_2 = 2\lambda \left[ \sum_{n=1}^{\infty} \ln(n) e^{-\lambda} \frac{\lambda^n}{n!} \right]\left[ \sum_{n=1}^{\infty} \ln\left(1+\frac{1}{n}\right) e^{-\lambda} \frac{\lambda^n}{n!} \right]$$
$$- 2\lambda \sum_{n=1}^{\infty} \ln(n)\ln(1+\frac{1}{n}) e^{-\lambda} \frac{\lambda^n}{n!} \qquad (B.9)$$

$$\gamma_3 = \left[ \sum_{n=1}^{\infty} \ln(n) e^{-\lambda} \frac{\lambda^n}{n!} \right]\left[ \sum_{n=1}^{\infty} \ln\left(1+\frac{1}{n}\right) e^{-\lambda} \frac{\lambda^n}{n!} \right]$$
$$+ \left[ \sum_{n=1}^{\infty} \ln(n) e^{-\lambda} \frac{\lambda^n}{n!} \right]^2 - \sum_{n=1}^{\infty} \ln^2(n) e^{-\lambda} \frac{\lambda^n}{n!} \qquad (B.10)$$

For the first item $\gamma_1$, we have

$$\sum_{n=1}^{\infty} \ln(n)\ln\left(1+\frac{1}{n}\right) e^{-\lambda} \frac{\lambda^n}{n!}$$
$$= \lambda \sum_{n=1}^{\infty} \frac{\ln(n)}{n} \ln\left(1+\frac{1}{n}\right) e^{-\lambda} \frac{\lambda^{n-1}}{(n-1)!}$$
$$\stackrel{n-1\to m}{=} \lambda \sum_{m=0}^{\infty} \frac{\ln(m+1)}{m+1} \ln\left(1+\frac{1}{m+1}\right) e^{-\lambda} \frac{\lambda^m}{m!} \qquad (B.11)$$
$$\stackrel{m\to n}{=} \lambda \sum_{n=1}^{\infty} \frac{\ln(n+1)}{n+1} \ln\left(1+\frac{1}{n+1}\right) e^{-\lambda} \frac{\lambda^n}{n!}$$

Hence,

$$\gamma_1 = \lambda \sum_{n=1}^{\infty} \frac{\ln(n+1)}{n+1} \ln\left(1+\frac{1}{n+1}\right) e^{-\lambda} \frac{\lambda^n}{n!} - \lambda \sum_{n=1}^{\infty} \ln^2\left(1+\frac{1}{n}\right) e^{-\lambda} \frac{\lambda^n}{n!} \quad (B.12)$$

Then, we can verify that only when $\lambda \leq 6$, $\gamma_1 < 0$. Since in experimental setups, $\lambda$ is always much larger than 6, we must have $\gamma_1 > 0$. For $\gamma_2$, according to Eq. (B.3), we have $\gamma_2 > 0$. For $\gamma_3$, the first derivative of $\gamma_3$ is larger than 0, and $\gamma_3 |_{n=1} = 0$ which means $\gamma_3 > 0$.

From Eq. (B.6), we have $\frac{dQ}{d\lambda} \geq \gamma_1 + \gamma_2 + \gamma_3 \geq 0$ which shows that $Q(\lambda)$ is monotonically increasing. In reality, the mean of collected x-ray photons, i.e., $\lambda$, is much larger than 100 and we have $Q(\lambda)|_{\lambda=100} > 0$. Thus,

$$Q(\lambda) = (1+\lambda)\left[\sum_{n=1}^{\infty} \ln(n) e^{-\lambda} \frac{\lambda^n}{n!}\right]^2 - \lambda \left[\sum_{n=1}^{\infty} \ln^2(n) e^{-\lambda} \frac{\lambda^n}{n!}\right] \geq 0 \quad (B.13)$$

That is,

$$\left[\sum_{n=1}^{\infty} \ln(n) e^{-\lambda} \frac{\lambda^n}{n!}\right]^2 \geq \lambda\left[\sum_{n=1}^{\infty} \ln^2(n) e^{-\lambda} \frac{\lambda^n}{n!} - \left(\sum_{n=1}^{\infty} \ln(n) e^{-\lambda} \frac{\lambda^n}{n!}\right)^2\right] \quad (B.14)$$

Hence, we obtain that

$$\Theta(0) = \frac{\sum_{n=1}^{\infty} \ln(n) e^{-\lambda} \frac{\lambda^n}{n!}}{\sqrt{\sum_{n=1}^{\infty} \ln^2(n) e^{-\lambda} \frac{\lambda^n}{n!} - [\sum_{n=1}^{\infty} \ln(n) e^{-\lambda} \frac{\lambda^n}{n!}]^2}} \geq \sqrt{\lambda} \quad (B.15)$$